\documentclass[12pt, twocolumn,showpacs,preprintnumbers,aps,prl,reprint,groupedaddress]{revtex4-1}

\usepackage{amssymb}
\usepackage{amsmath}
\usepackage{subfigure}
\usepackage{graphicx}
\usepackage{color}

\begin{document}

\preprint{}

\title{Vorticity, Defects and Correlations in Active Turbulence}


\author{Sumesh P. Thampi}
\affiliation{The Rudolf Peierls Centre for Theoretical Physics, 1 Keble Road, Oxford, OX1 3NP, UK}

\author{Ramin Golestanian}
\affiliation{The Rudolf Peierls Centre for Theoretical Physics, 1 Keble Road, Oxford, OX1 3NP, UK}

\author{Julia M. Yeomans}
\affiliation{The Rudolf Peierls Centre for Theoretical Physics, 1 Keble Road, Oxford, OX1 3NP, UK}
\email[]{j.yeomans1@physics.ox.ac.uk}
\homepage[]{http://www-thphys.physics.ox.ac.uk/people/JuliaYeomans/}




\begin{abstract}
We describe a numerical investigation of a continuum model of an active nematic, concentrating on the regime of active turbulence. Results are presented for the  effect of three  parameters, activity,  elastic constant and rotational diffusion constant, on the order parameter and flow fields. Defects and distortions in the director field act as sources of vorticity, and thus vorticity is strongly correlated to the director field. In particular the characteristic length of decay of vorticity and order parameter correlations is controlled by the defect density. By contrast the decay of velocity correlations is determined by a balance between activity and dissipation.  We highlight the role of microscopic flow generation mechanisms in determining the flow patterns and characteristic scales of active turbulence and contrast the behaviour of extensile and contractile active nematics.
\end{abstract}



\maketitle

\section{Introduction}

\begin{figure*}
\includegraphics[width=\linewidth]{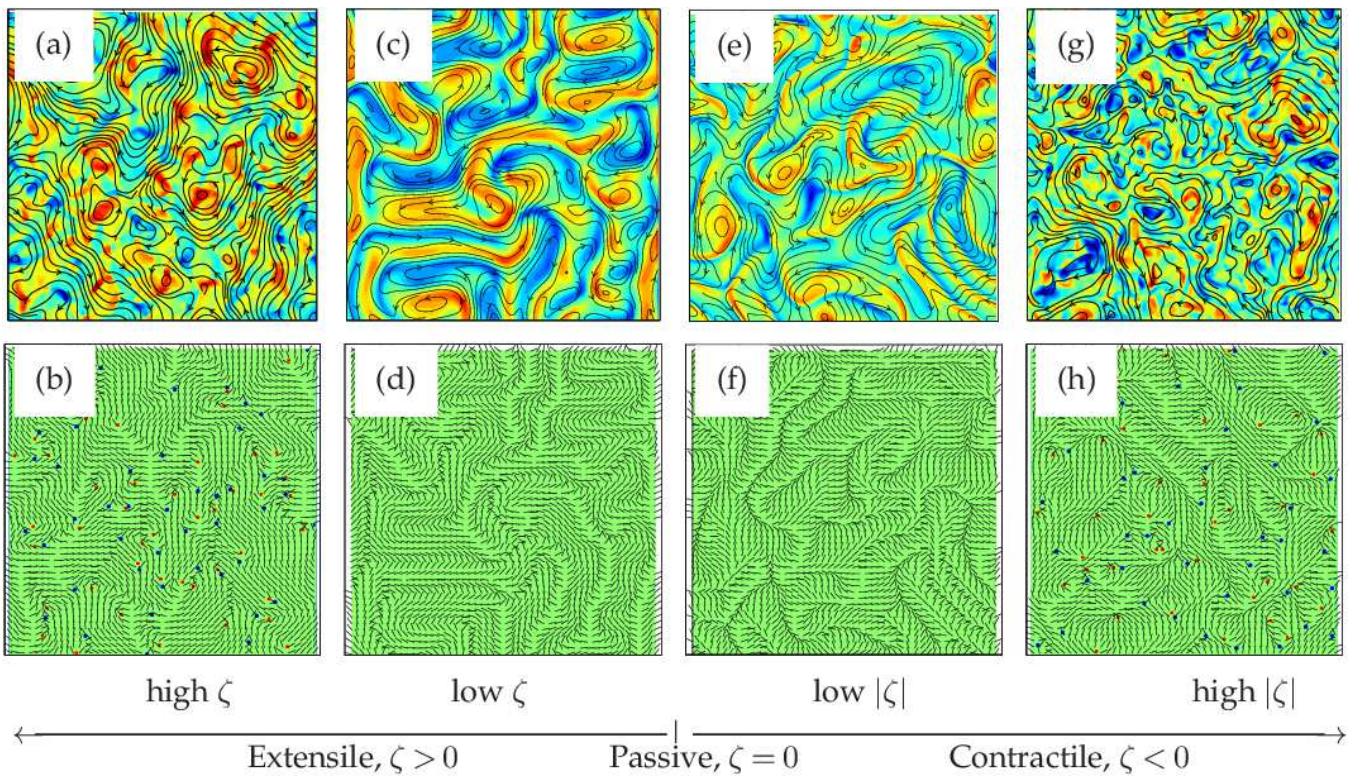}
\caption{Fully developed active turbulent flow and director patterns from simulations of active nematics comparing an extensile and contractile system for high and low values of activity coefficient, $\zeta$. Streamlines and vorticity field, with red and blue colouring corresponding to high $+$ve and $-$ve vorticity respectively are shown for (a) an extensile system with large activity ($\zeta=0.0125$), (c) an extensile system with small activity ($\zeta=0.001$), (e) a contractile system with small activity ($\zeta=-0.001$) and (g) a contractile system with large activity ($\zeta=-0.0125$). Vorticity is scaled with its absolute maximum value in each panel. The corresponding director field and topological defects ($+{1}/{2}$, red; $-{1}/{2}$, blue), if present, are shown in panels (b), (d), (f), (h). Panels (b) and (h) show the left bottom quarter of the domain of (a) and (g) respectively.}
\label{fig:fullfledge}
\end{figure*}

Active systems produce their own energy and thus operate out of thermodynamic equilibrium. Examples include suspensions of microtubules and molecular motors, cellular and bacterial suspensions, vibrating granular rods, flocks of birds, and schools of fish \cite{dogic2012, Sriram2010, Marchetti2013, Jorn2013,Chate2012,Cavagna2010}.  In all these systems, the motion is generated at the level of individual constituents but energetic, hydrodynamic or intelligent interactions then result in a collective dynamical behaviour. In particular
the spontaneous motion generated by a wide variety of dense active systems is highly chaotic, consisting of fluid jets and swirls (see fig. \ref{fig:fullfledge}). The flow pattern visually resembles that of  high $Re$ number turbulence and hence this state is often  termed  `active turbulence'  \cite{julia2012}.  It is not clear whether the similarities in the turbulent states observed in active systems across a wide range of length scales are superficial, or represent underlying universal properties. 


Many active systems have nematic symmetry and exhibit orientational order,  and concepts from  liquid crystal physics have been used to describe them. The equations of nematohydrodynamics have proved useful in predicting the occurrence of hydrodynamic instabilities \cite{Sriram2002, Joanny2005, Madan2007, Davide2007, Scott2009} and in simulating the turbulent flow behaviour \cite{Mahadevan2011, ourprl2013, ourepl2014}. There is also increasing evidence that topological defects and their dynamics play a role in active turbulence  \cite{Mahadevan2011, Giomi2013, ourprl2013}.  Recent experiments using microtubule bundles and kinesin molecular motors have demonstrated the presence of topological defects in active systems \cite{dogic2012}. Similar defects have also been observed in experiments on living liquid crystals \cite{Igor2013} and vibrating granular rods \cite{Narayan2007} and in the simulations of self-propelling ellipsoidal particles \cite{Shi2013}.

In this paper we report extensive simulations of active nematics to investigate and characterise the flow field and the order parameter field, and the coupling between them. We find that two length scales are relevant. For extensile systems scaling arguments suggest that the length describing the decay of velocity correlations results from a balance between activity and dissipation. Correlation functions of the director and vorticity fields have very similar forms which correspond to a different length scale, the defect spacing. We report different behviour for extensile and contractile systems, showing the sensitivity of the active state to details of the coupling between the director and flow fields.

In section 2 we summarise the theory of nematohydrodynamics and show that it can be extended to the active case by including a simple, additional, dipolar source term in the stress. Section 3 gives details of the numerical approaches used to solve the equations of motion. In Section 4 we give a qualitative account of the characteristic features of active turbulence in extensile and contractile systems explaining, in particular, how defects are continually created, and subsequently annihilated, to give a steady state with a finite defect density. Section 5 presents extensive quantitative results showing how the magnitude and correlations of velocity, vorticity and the order parameter field depend on model parameters. We also measure the number of defects and their rate of creation and annihilation. In section 6 we show that an analysis of the defect density and dynamics can be used to relate the different scalings found in the numerics. We also list open questions that warrant attention in the future.


\section{Equations of Motion}

We first describe the hydrodynamic equations of motion of an active nematic system. This may be a dense suspension of microtubule bundles and molecular motors \cite{dogic2012}, or a dense suspension of bacteria \cite{julia2012}, or a mixture of bacteria and liquid crystal molecules \cite{Igor2013}. Active systems with nematic symmetry, and density and momentum as conserved variables, can be modelled using the standard equations of nematohydrodynamics, modified to incorporate an active term which produces flows in response to gradients in orientational order \cite{Sriram2002, Marchetti2013}. The magnitude and direction of the orientational order is described by a second rank tensor  $\mathbf{Q}$ which is symmetric ($Q_{ij}=Q_{ji}$) and traceless ($Q_{ii}=0$). For uniaxial nematics $\mathbf{Q} = \frac{q}{2} ( 3\mathbf{nn} - \mathbf{I})$ where $q$ is the largest eigenvalue of $\mathbf{Q}$ and is a measure of nematic degree of order, $\mathbf{n}$ is the director field and $\mathbf{I}$ is the Identity tensor.

The nematohydrodynamic equations can be derived using a Poisson bracket formalism or from linear irreversible thermodynamics with phenomenological models assumed for the relation between fields and currents. The resulting equations governing the evolution of the order parameter $\mathbf{Q}$ and momentum $\rho \mathbf{u}$ 
are \cite{Berisbook, DeGennesBook},
\begin{align}
(\partial_t + u_k \partial_k) Q_{ij} - S_{ij} &= \Gamma H_{ij},
\label{eqn:lc}\\
\rho (\partial_t + u_k \partial_k) u_i &= \partial_j \Pi_{ij};~~~~\partial_i u_i = 0.
\label{eqn:ns}
\end{align}
The generalised nonlinear advection term in eq.~(\ref{eqn:lc}) is
\begin{align}
S_{ij} = &(\lambda E_{ik} + \Omega_{ik})(Q_{kj} + \delta_{kj}/3) + (Q_{ik} + \delta_{ik}/3)\nonumber\\
&(\lambda E_{kj} - \Omega_{kj}) - 2 \lambda (Q_{ij} + \delta_{ij}/3)(Q_{kl}\partial_k u_l)\nonumber
\end{align}
where $E_{ij} = (\partial_i u_j + \partial_j u_i)/2$ is the strain rate tensor and $\Omega_{ij} = (\partial_j u_i - \partial_i u_j)/2$ is the vorticity tensor. These characterise the deformational and the rotational components of the flow field respectively. We also define the vorticity, $\boldsymbol{\omega} = \nabla \times \mathbf{u}$ which measures the rotation of the fluid elements. It is related to the vorticity tensor by $\epsilon_{ijk}\omega_k = -2\Omega_{ij}$ where $\boldsymbol{\epsilon}$ is the Levi-Civita symbol. We will use vorticity extensively to characterise the flow field in this paper.

The alignment parameter $\lambda$ controls the degree of coupling between the orientation field and velocity gradients \cite{DeGennesBook, Scott2009}. Mathematically $\lambda$ determines the character of the objective time derivative of $\mathbf{Q}$. The values $\lambda=1$ and $\lambda=-1$  correspond to an upper and lower convected derivative and $\lambda=0$ corresponds to the corotational derivative which is obtained as an average of the lower and upper convected derivatives.
On mapping to the Leslie Erickson equation which describes the liquid crystal dynamics in terms of the director field $\mathbf{n}$ only, $(3q+4)\lambda/9q$ is the flow alignment parameter \cite{Davide2007} which determines whether the liquid crystal molecules (or here, the active constituents) align or tumble in a shear flow.

In linear approximation, the relaxation of $\mathbf{Q}$ is driven by the molecular field
\begin{align}
H_{ij} = -\frac{\delta \mathcal{F}}{ \delta Q_{ij}} + \frac{\delta_{ij}}{3} {\rm Tr} \frac{\delta \mathcal{F}}{ \delta Q_{kl}},
\end{align}
 defined as the variational derivative of a free energy functional. $\mathbf{H}$ is also symmetric and traceless. The phenomenological constant of proportionality, which corresponds physically to the rotational diffusion coefficient, is $\Gamma$. The Helmholtz free energy,
\begin{align}
\mathcal{F} &= \frac{A}{2} Q_{ij} Q_{ji} +\frac{ B}{3} Q_{ij} Q_{jk} Q_{ki} +\frac{C}{4} (Q_{ij} Q_{ji})^2\nonumber\\
 &+ \frac{K}{2} (\partial_k Q_{ij})^2,
\end{align}
has a bulk contribution which is a truncated polynomial expansion in invariants of $\mathbf{Q}$ \cite{Berisbook, DeGennesBook} and  a term arising from the gradients in $\mathbf{Q}$. This free energy expression is general in that it depends on gradients of the scalar order parameter in addition to the gradients in the director orientation. $A, B$ and $C$ are phenomenological coefficients which are functions of concentration and temperature. Here we use a single elastic constant ($K$) approximation.

In eq. (\ref{eqn:ns}) the total stress generating the hydrodynamics includes:
\begin{enumerate}
\item The viscous stress, 
\begin{equation}
\Pi_{ij}^{viscous} = 2 \mu E_{ij} ,
\label{eqn:viscstress}
\end{equation}
where $\mu$ is the Newtonian viscosity of the suspension.
\item  The orientational elastic stresses of a passive liquid crystal,
\begin{align}
&\Pi_{ij}^{passive}=-P\delta_{ij} + 2 \lambda(Q_{ij} + \delta_{ij}/3) (Q_{kl} H_{lk})\nonumber\\
&-\lambda H_{ik} (Q_{kj} + \delta_{kj}/3) - \lambda (Q_{ik} + \delta_{ik}/3) H_{kj}\nonumber\\
&-\partial_i Q_{kl} \frac{\delta \mathcal{F}}{\delta \partial_j Q_{lk}} + Q_{ik}H_{kj} - H_{ik} Q_{kj}.
\label{eqn:passstress}
\end{align}
As a result of this term the orientational order can affect the flow field, an effect often referred to as the 'back-flow'. The back-flow can modify the director dynamics, for example accelerating the defect annihilation process \cite{Julia2002} and affecting the motion of active defects \cite{Giomi2013}. The term $P = \rho T - \frac{K}{2} (\partial_k Q_{ij})^2$ in eq.~(\ref{eqn:passstress}) is the modified bulk pressure.

\item The active stress, 
\begin{equation}
\Pi_{ij}^{active} = -\zeta Q_{ij} ,
\label{eqn:actistress}
\end{equation}
is introduced in \cite{Sriram2002}. This corresponds to a force dipole, which is the leading order contribution of activity to the stress. The strength of this term is prescribed by the activity coefficient $\zeta$, which is +ve for extensile and -ve for contractile systems. Gradients in $\mathbf{Q}$ thus produce a flow field, in addition that due to the passive stress terms, which is the source of the hydrodynamic instability in active nematics.
\end{enumerate}

More details about this model and its application to passive and active systems can be found in  \cite{Berisbook, DeGennesBook, Denniston2001, Denniston2004, Davide2007, Cates2008, Orlandini2008, Henrich2010, Suzanne2011, Miha2013}. The nematic state has been shown to be unstable to flow induced by the activity \cite{Sriram2002} and the perturbations of the order parameter field that drive the instability have been studied \cite{Madan2007, Scott2009}. The onset of activity driven flow in a 1-D channel was analysed in \cite{Joanny2005} and the flow patterns generated by the activity were studied numerically \cite{Davide2007, Miha2013, Cates2009}. The rheology of active materials has also been addressed \cite{Davide2007B, Suzanne2011}. The role of topological defects in active systems has also attracted much attention recently \cite{Mahadevan2011, ourprl2013, Giomi2013}.

\section{Simulations}

The governing equations (\ref{eqn:lc}) and (\ref{eqn:ns}) form a coupled system, and we use a hybrid numerical method to solve them \cite{Davide2007, Suzanne2011, Miha2013}.  Instead of directly solving the momentum equation a lattice Boltzmann method is used to determine the hydrodynamics. For the order parameter evolution equation, we employ a method of lines where all spatial derivatives are calculated using a finite difference approach. The resulting system of ordinary differential equations in time is solved using Euler's integration method \cite{Davide2007B, Henrich2010}. At every time step the order parameter field obtained from the method of lines is an input to the lattice Boltzmann simulations and then the velocity field obtained from the lattice Boltzmann is used in the next time step of the method of lines for updating $\mathbf{Q}$. The computations are performed until a statistical steady state is reached. 

For the lattice Boltzmann simulations, we use a $D3Q15$ model where the space and time discretizations are chosen as unity.  Simulations are performed on a two dimensional domain of size $400 \times 400$ with periodic boundary conditions in all directions. The parameters used are $\Gamma=0.34$, $A=0.0$, $B=-0.3$, $C=0.3$, $K=0.02$, $\mu=2/3$ unless mentioned otherwise. We choose $\lambda=0.7$ corresponding to aligning flows \cite{DeGennesBook, Scott2009}. All quantities are reported in lattice Boltzmann units and depending upon the system of interest, eg microtubule or bacterial suspensions, appropriate conversion to physical units may be performed \cite{Cates2008, Henrich2010, ourprl2013}. 

\begin{figure*}
\includegraphics[width=0.9\linewidth]{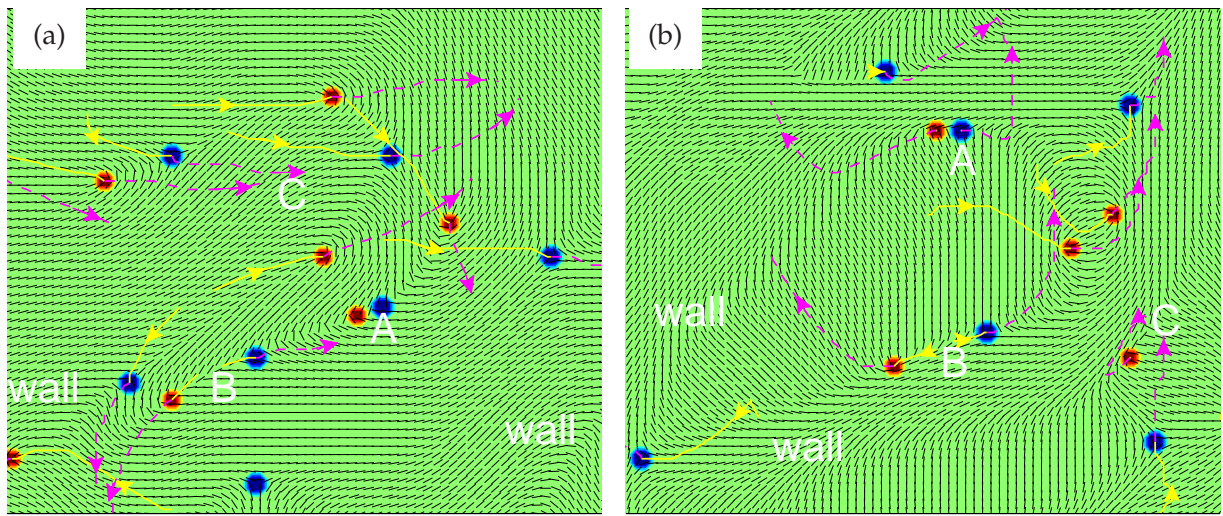}
\caption{Snap shots of defect dynamics in (a) an extensile system and (b) a contractile system. Some walls are indicated and $+\frac{1}{2}$ and $-\frac{1}{2}$ defects are highlighted as red and blue respectively. The past and future trajectories of the defects are shown with continuous (yellow) and dashed (magenta) lines respectively. An example of (i) defect creation is labelled `A', (ii) defects moving along the walls is labelled `B' and (iii) defect annihilation is labelled `C'.
 }
\label{fig:defdyn}
\end{figure*}

\section{Active turbulence in extensile and contractile systems}
\label{sec:acttur}

Here we give a qualitative description of active turbulence, before presenting and discussing quantitative data for the turbulent state.

Fig. \ref{fig:fullfledge} shows typical snapshots of fully developed turbulent patterns for a two dimensional active nematic comparing extensile and contractile stresses and high and low values of the activity. For each set of parameters, the flow field is represented by firstly the stream lines which reveal the circulating patterns and secondly the colour-shaded vorticity field which reveals the direction of rotation. The flow fields are highly disordered in both the extensile and contractile cases, with characteristic swirling patterns. The corresponding director fields are shown in the second row. These also exhibit a high degree of disorder although, for the parameters we use, the corresponding passive liquid crystal is nematic (not shown). 

Due to active stress, flow is generated in response to distortions in the order parameter field. The flow further disturbs the director field thus resulting in active turbulence. Since the dominant instability and hence the mechanism for flow generation is bend distortions of the director field for extensile systems and splay distortions for contractile systems \cite{Madan2007, Scott2009}, both the flow and director fields show qualitative differences in their structures. 

In the transition to active turbulence, active nematic regions develop \textit{walls}, which are elongated distortions in the director field.
Correspondingly, the flow field also exhibits elongated structures of vorticity (fig:\ref{fig:fullfledge}(c), \ref{fig:fullfledge}(e)). We shall call this active turbulence at `low activity'.  In low activity regime, the walls persist but are continuously advected and deformed due to flow and flow gradients. 

In contrast, at `high activity' the walls are strongly deformed by the flow generated by the activity. They are unstable and give rise to the formation of  pairs of oppositely charged  $\pm \frac{1}{2}$ defects. The defect formation is driven by both  elasticity and flow \cite{ourepl2014}. These topological defects of charge $\pm\frac{1}{2}$ are distinguished by red and blue colouring in the director field in fig.~\ref{fig:fullfledge} (second row). As a result of the stronger velocity fields and the defect formation the vorticity structures in the flow field are also more isotropic (fig:\ref{fig:fullfledge}(a), \ref{fig:fullfledge}(g)) compared to those in the low activity regime. Topological defects in active systems have been observed in a variety of experimental systems and simulation studies \cite{dogic2012, Igor2013, Shi2013}.

Details of defect dynamics for an extensile and a contractile system are shown in fig.~\ref{fig:defdyn}.
Consider the extensile system first (fig.~\ref{fig:defdyn}(a)).  Some of the walls are indicated in the figure.  The $+\frac{1}{2}$ and $-\frac{1}{2}$ defects are highlighted with red and blue colour and their trajectories are also shown. Yellow continuous lines are their past, and magenta dashed lines their future paths. It is apparent that the defects are formed in walls. They preferentially move along the wall releasing the associated bend free energy. 
During this motion, defects encounter oppositely charged defects and annihilate, thus reinstating nematic regions. All these events are illustrated in fig.~\ref{fig:defdyn}(a) e.g.: a creation event is labelled as `A', defects moving along the walls as `B' and a defect annihilation event as `C'.  A similar snapshot of defect dynamics for a contractile system is shown in fig.~\ref{fig:defdyn}(b). Since the dominant mode of instability for a contractile nematic is splay deformations in the director field, the walls appear different in this case. As can be seen in fig.~\ref{fig:defdyn}(b) they form the borders of nematic regions of different mean orientations. Again defects are formed in the wall, and preferentially move along the wall. As in the extensile case, they then encounter defects of opposite charge and annihilate re-establishing nematic regions. 

Since walls tend to be very long, more than one pair of defects may be produced from a single wall, at the same or different times. Defects tend to move along walls and therefore are more likely to encounter defects of opposite charge formed in the same wall and quickly annihilate each other. Some of the defects escape from the walls. The search for a defect of opposite sign is then two dimensional and the defects'  lifetimes are considerably longer.


\begin{figure*}
\includegraphics[width=0.9\linewidth]{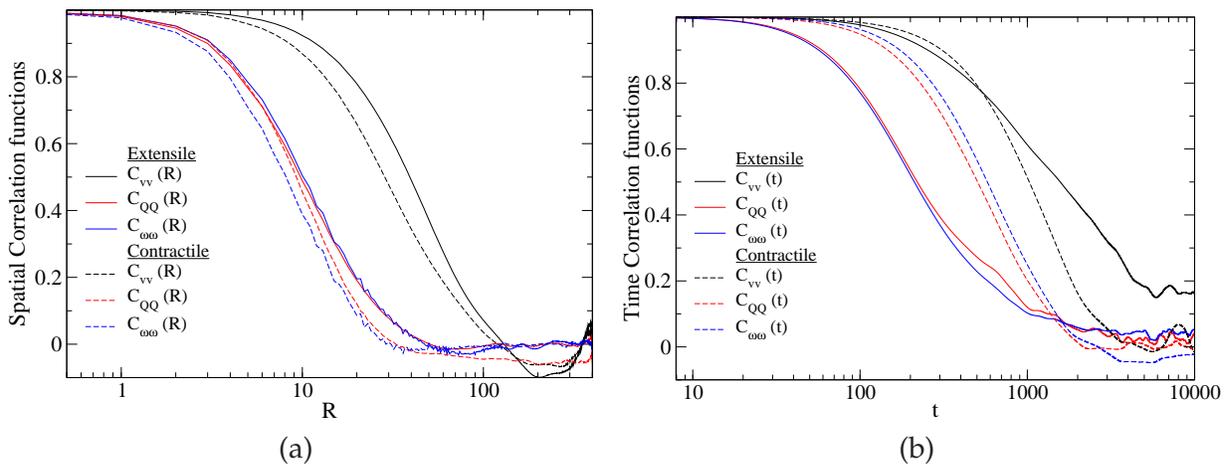}
\caption{Correlation functions of velocity, $\mathbf{v}$, order parameter, $\mathbf{Q}$, and vorticity $\boldsymbol{\omega}$ fields are shown for both extensile (continuous line) and contractile (dashed line) systems. Panel (a) shows spatial and panel (b) shows time correlation functions. All functions $<\cdot>$ are normalised. Also$<\mathbf{Q:Q}>$ is scaled in y to lie between $0$ to $1$. The x-axis is not scaled. $\Gamma=1$ is used in these simulations. The vorticity and order parameter correlations have a close association with similar decays and the same characteristic length and time scales of decay for both extensile and contractile systems. The velocity correlation functions correspond to different length and time scales.}
\label{fig:vvnnww}
\end{figure*}

\section{Results}
\label{sec:results}

\subsection{Correlation functions}
\label{sec:corr}

Vorticity
 quantifies the local rotation of the fluid elements in a flow field. Many different active systems exhibit continually changing patterns of vorticity \cite{julia2012, Jorn2013, Lenepreprint, Miha2013, Goldstein2012} and it was suggested in \cite{julia2012, Jorn2013} that vorticity might be helpful in characterising active turbulence.
Indeed vorticity is a fundamental concept in inertial turbulence because regions of high vorticity are characteristic features of turbulent flows. Here we argue that it is also a key quantity in active turbulence, but for a different reason. 

In order to investigate and quantify the structure of the flow and ordering, we plot spatial correlation functions of the velocity, the vorticity and the order parameter field in fig.~\ref{fig:vvnnww}(a) and temporal correlations functions of the same quantities in fig.~\ref{fig:vvnnww}(b). To allow an easier comparison the correlation functions are normalised to unity (i.e, $C_{vv}(R) = \langle \mathbf{v} (R,t)\cdot\mathbf{v}(0,t) \rangle / \langle\mathbf{v} (0,t)^2\rangle$, $C_{\omega\omega}(R) = \langle \boldsymbol{\omega} (R,t)\cdot\boldsymbol{\omega}(0,t) \rangle / \langle\boldsymbol{\omega} (0,t)^2\rangle$) and the order parameter correlation functions are further scaled to lie between 0 and 1 (i.e,  $C_{QQ}(R) = \Big(\langle\mathbf{Q} (R,t):\mathbf{Q}(0,t)\rangle-\langle\mathbf{Q} (\infty,t):\mathbf{Q}(0,t)\rangle\Big)/\left(\langle \mathbf{Q} (0,t)^2\rangle-\langle\mathbf{Q} (\infty,t):\mathbf{Q}(0,t)\rangle\right)$, etc.). There is no scaling of  the x-axis in any plot.

 For each correlation function extensile and contractile systems are compared. (Here, and throughout the paper, we use continuous (dashed) lines to depict results for extensile (contractile) systems.)
The only difference between the parameters for the extensile and contractile simulations  is the sign of the activity coefficient $\zeta$. As is clear from the figure, extensile and contractile systems exhibit different characteristic length and time scales of decay for all the quantities measured. This is not surprising considering that the dominant mechanisms for flow generation vary between the two cases.

Note, however, that the  director field and vorticity field  have a very similar dependence on $R$ and $t$, whereas the behaviour of the velocity correlations is quite different. This close association between the order parameter and the vorticity is seen for both the extensile and the contractile simulations. This suggests that vorticity is a fundamental quantity in active turbulence, which is
closely related to the microstructure of the system. We shall present more evidence for this assertion in section \ref{sec:results}\ref{sec:effects} when we find that the characteristic lengths associated with the vorticity field and the order parameter field scale in the same fashion with several system parameters.

\begin{figure*}
\includegraphics{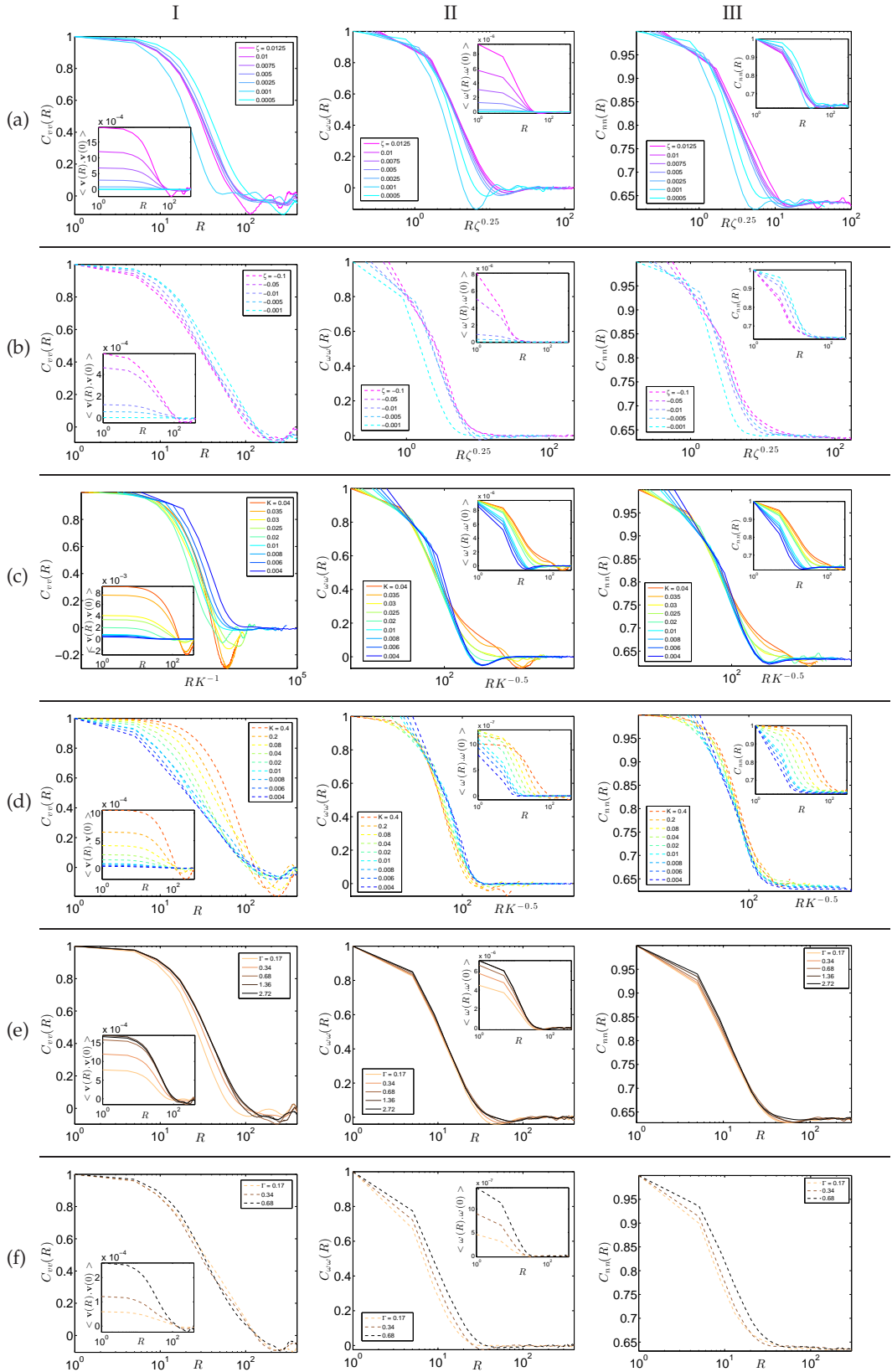}
\caption{Spatial correlations of I velocity, II vorticity,  and III director fields for different values of activity coefficients of (a) an extensile and (b) a contractile system; elastic constants for (c) an extensile and (d) a contractile system; rotational diffusion constant for (e) an extensile and (f) contractile system. $\zeta=0.01$ is used in the simulations for rows (e) and (f). In each case the correlation functions are normalised and the length $R$ is scaled to give an optimum data collapse. The inset figures show the unscaled functions.}
\label{fig:zetaKdep}
\end{figure*}

\begin{figure*}
\includegraphics[width=0.8\linewidth]{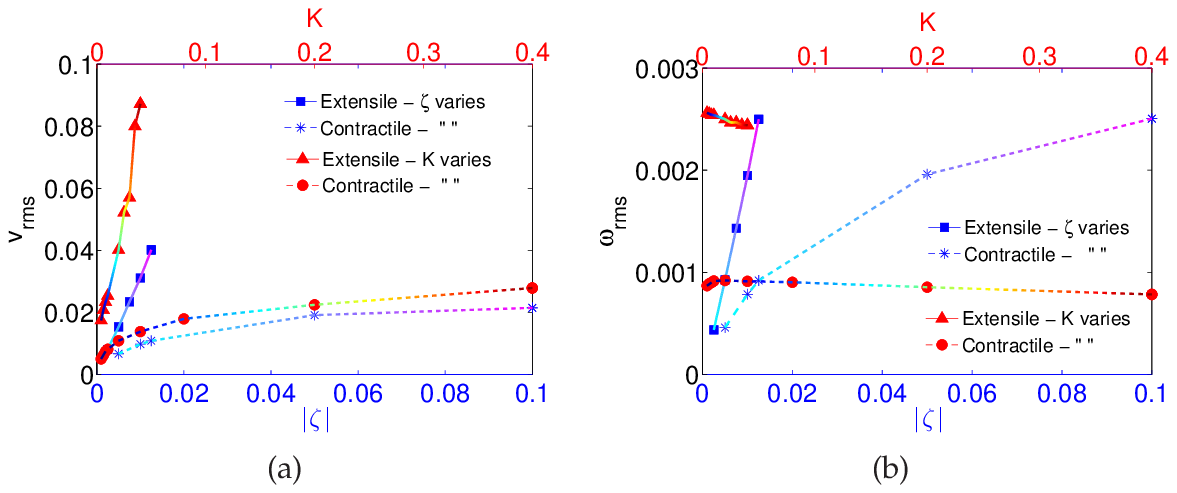}
\caption{(a) RMS velocity and (b) RMS vorticity of the active turbulent flow field as a function of the activity coefficient and the elastic constant for both extensile and contractile systems.}
\label{fig:vrmspara}
\end{figure*}

The connection between vorticity and local order also follows from the governing equations of motion.
Generally active systems such as suspensions of microtubules and molecular motors or bacteria operate at small, often negligible, Reynolds number indicating that these are inertia-less systems and that diffusion of momentum is instantaneous. In this limit, the Navier - Stokes equations considerably simplify to obtain $\partial_j \Pi_{ij} = 0$. Taking the curl of this equation and substituting the expressions~(\ref{eqn:viscstress}-\ref{eqn:actistress}) for the viscous, passive and active stresses, gives
\begin{equation}
\mu \nabla^2 \boldsymbol{\omega} + \zeta \nabla \times \nabla \cdot \mathbf{Q} + \nabla \times \nabla \cdot \boldsymbol{\Pi}^{\textnormal{Passive}}(\mathbf{Q})= 0 .
\label{eqn:curl}
\end{equation}
This indicates that gradients in $\mathbf{Q}$ act as sources of vorticity which diffuse in the system generating vorticity patterns and thus it is reasonable to expect that vorticity and order parameter fields may show a close association in their behaviours.

\subsection{Dependence on parameters}
\label{sec:effects}

In the last section we showed that the decay of the normalised correlation functions of the director field and the vorticity are governed by the same length scale, but that the velocity correlation function decays over a longer length scale. We  now show how the correlation functions depend on three of the most important parameters of the system, the activity coefficient $\zeta$, the elastic constant $K$ and the rotational diffusion constant $\Gamma$ . The results are collected in fig.~\ref{fig:zetaKdep}. The three columns of the figure show $C_{vv}(R)$, $C_{\omega\omega}(R)$ and $C_{nn}(R)$. (Note that we plot the director correlation function, $C_{nn}(R) = \langle \mathbf{n} (R,t)\cdot\mathbf{n}(0,t) \rangle$ instead of that of
 the order parameter $C_{QQ}(R)$. Both behave in the same fashion.) 

The first two rows of the figure show how the correlation functions depend on activity $\zeta$ for extensile and contractile nematics respectively. Rows (c) \& (d) show results for the dependence on the elastic constant $K$ and the final rows on the diffusion constant $\Gamma$. In each case the horizontal distance axis is scaled with the relevant parameter to achieve optimal data collapse. The inset in each figure shows the unscaled data. Since the dominant hydrodynamic instability in contractile systems is to splay configurations which yield lower velocities than the bend configurations which occur more frequently in extensile nematics\cite{Scott2009} we have been able to cover a larger range of activity and elastic coefficients in the contractile systems.

The optimum scalings with respect to different parameters are based on visual observation of the best collapse and the accuracy of the scaling exponent  $\sim \;\; \pm 0.05$. A nearest rational number is always chosen for brevity of discussion.

The resemblance between the second and third columns of fig.~\ref{fig:zetaKdep} is immediately apparent, confirming the similar behaviour of the vorticity and the order parameter correlations. As the latter two quantities scale in the same way we shall just compare the scaling of the velocity and director correlations (columns I and III) in the following. The characteristic length for the director field is $\ell_{\rm n} \sim \sqrt{K}/\zeta^{1/4}$ for both extensile and contractile systems. The length scale $\ell_{\rm vel}$  governing decay of velocity correlations is independent of $\zeta$ for both extensile and contractile nematics  \cite{Aranson2012, Jorn2013, ourprl2013}. It scales as $\sim K$ for extensile systems but 
we have not found any obvious scaling behaviour for the contractile ones.
 None of the correlation functions depend on $\Gamma$ for either extensile or contractile systems.

Note that in both extensile and contractile systems, activity coefficients $\lesssim 0.001$ correspond to the low activity regime. In this regime, no topological defects form yielding elongational flow patterns that are visually different to the high activity patterns (fig: \ref{fig:fullfledge}). Thus the scalings are expected to be different in the low activity regime as is indeed seen in fig.~\ref{fig:zetaKdep}. Similarly at sufficiently large values of $\Gamma$, the order parameter field relaxes fast and very few defects have time to form in the walls. This again results in elongated structures of both director and flow fields similar to those formed in the low activity regime. Due to this, only a small range of $\Gamma$ could be scanned in contractile systems.

We also plot the root mean square (RMS) value of the velocity and vorticity as a function of $\zeta$ and $K$ in fig.~\ref{fig:vrmspara}, for both extensile and contractile systems. The RMS velocity increases with $\zeta$ and $K$, with extensile systems showing a stronger dependence than contractile systems. The RMS vorticity increases with $\zeta$ but is insensitive to changes in $K$ for both extensile and contractile systems.

\begin{figure*}
\includegraphics[width=\linewidth]{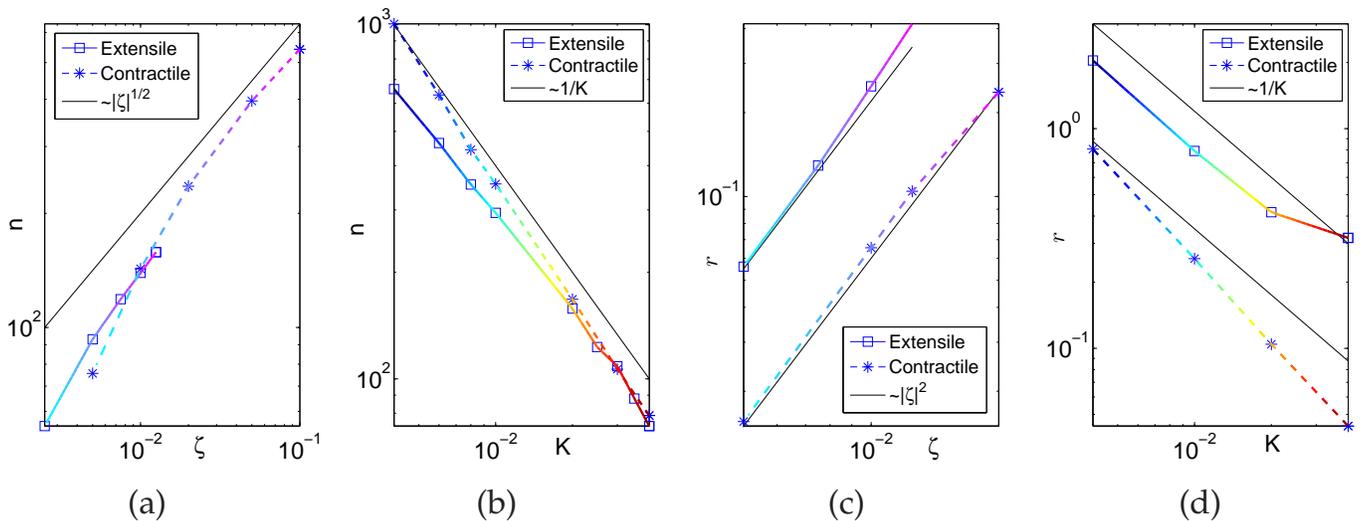}
\caption{(a) Number of $\pm \frac{1}{2}$ defects $n\sim \sqrt{\zeta}$ and (b) $n\sim 1/K$ for both extensile and contractile systems. This agrees with the scalings observed for director correlation function. (c) Rate of creation and annihilation of defect pairs $r \sim |\zeta|^2$ and (d) $r \sim1/K$ in both extensile and contractile systems.}
\label{fig:defmea}
\end{figure*}


Since we saw in section \ref{sec:acttur} that topological defects appear to be important in controlling active turbulence we next present numerical results for 
the number of defects and their rate of formation and destruction. Since we find only a weak dependence on $\Gamma$ in these and previous measurements, we exclude this parameter from further discussion. In fig.~\ref{fig:defmea}(a)-(b) we plot the number of defects as a function of activity and elasticity for extensile and contractile systems finding
\begin{equation}
n \sim \sqrt{\zeta}/K
\end{equation}
in both cases. 
At steady state the rate of creation and the rate of annihilation of defects are equal.  We report this rate in fig.~\ref{fig:defmea}(c)-(d) giving a scaling with activity and elasticity as $r \sim {\zeta^2}/{K}$.\\


\section{Discussion}
We now discuss the scaling behaviour of the different dynamical variables and suggest a framework in which they can be connected by considering the motion of defects. We will first consider extensile nematics, and argue that $\ell_{\rm vel}$, the length scale of the velocity field, is controlled by a balance between activity and dissipation, whereas $\ell_{\rm n}$, the length scale of the order and the vorticity, is set by the defect spacing.

Generalising an argument given in \cite{ourprl2013}, we consider the rate of energy input into the system by the active stress and assume that this balances the viscous dissipation over a domain of size $\ell_{\rm vel}$. Using eqn.~(\ref{eqn:ns}), we find a scaling for the magnitude of the velocity
\begin{equation}
v \sim \zeta \ell_{\rm vel} q/ \mu.
\label{eqn:flowv}
\end{equation}
The results presented in fig.~\ref{fig:defmea} show that the rate of creation of defects in active turbulence is $\sim \zeta^2/K$.  We estimate the rate of annihilation of defect pairs of opposite charge using simple kinetic theory arguments. If $\sigma$ is the scattering cross section for collisional events leading to annihilation and $n$ is the number density of defects, then $1/n \sigma$ is the mean free path and the rate of annihilation $\sim   \sigma v  n^2$ where we have made the implicit assumption that the defect and the fluid velocities scale in the same way.
At steady state the rate of generation and the rate of annihilation  of defects are equal and thus we obtain
\begin{equation}
\frac{\zeta^2 }{K} \sim \frac{\sigma \zeta \ell_{\rm vel} q n^2}{\mu}\;\;\;\; \Rightarrow \;\;\;\;  \ell_{\rm vel} \sim \frac{\zeta}{K n^2},
\end{equation}
assuming that $\sigma$, $\mu$ and $q$ are independent of $K$ and $\zeta$.
The data in fig. \ref{fig:defmea}(a)-(b) indicates that $n \sim \sqrt{\zeta} /K$. Therefore we expect
\begin{equation}
\ell_{\rm vel} \sim K 
\label{eqn:lscale}
\end{equation}
in agreement with the numerical results in 
figs.~4(a) and 4(c). Substituting the expression~(\ref{eqn:lscale}) back into eqn.~(\ref{eqn:flowv}), gives $v \sim \zeta K$, in agreement with the results for the RMS velocity reported in fig.~\ref{fig:vrmspara}(a).

A second characteristic length in the system is that which controls the decay of both the order parameter field and the vorticity $\ell_{\rm n} \sim \sqrt{K}/ \zeta^{1/4}$. Our results show that the number of defects $\sim \zeta^{1/2}/ K \sim \ell_{\rm n}^{-2}$, suggesting that the decay of both the director field and the vorticity is controlled by the distance between defects, and hence that the defects are acting as sources of vorticity. The magnitude of the vorticity, however, would be expected to scale as a derivative of the velocity, $\sim v/\ell_{\rm vel} \sim \zeta$, as confirmed by the data in fig.~\ref{fig:vrmspara}(b). It is interesting to note that the length scale governing the hydrodynamic instability  $\sim \sqrt{K/\zeta}$ appears to be unrelated to the scales $\ell_{\rm vel}$ and $\ell_{\rm n}$ measured in the fully developed turbulent state \cite{ourepl2014}. 

Thus a simple scaling picture agrees well with the numerical results for extensile nematics. This is not the case for contractile systems where, for example, the velocity-velocity correlation function does not show any simple scaling with $K$. This may be because flow in contractile systems is generated by both topological defects which comprise predominantly bend distortions and by the splay deformations which arise from hydrodynamic instabilities. It may also be due to the shear thickening known to occur in contractile systems which will introduce a dependence of $\mu$ on activity.

Our numerical results suggest many avenues for further research. For example, we lack a theory of why defects are created at a rate ${\zeta^2 }/{K}$ or for the cross section for defect annihilation. Moreover the mechanisms whereby defects and distortions create vorticity in active systems remain to be better understood. It will also be interesting to obtain similar numerical results for other systems which show active turbulence, for examples those with polar symmetry or damped hydrodynamics, to identify which properties of the turbulent state are universal. The extent to which the equations~(\ref{eqn:lc}-\ref{eqn:actistress}) are appropriate to describe experiments on microtubules \cite{dogic2012}, which are long elastic filaments, remains an open question.

In conclusion, we have performed  numerical simulations of  active turbulence in extensile and contractile nematics. We find that the same length scale is associated with correlations of the order parameter field and of the vorticity field. This suggests that defects and distortions are acting as sources of vorticity, in agreement with visual inspection of the simulation results, and provides evidence that vorticity is an important quantity in describing active turbulence \cite{julia2012}.

\section*{Acknowledgment}
The authors thank Z. Dogic, D. Pushkin and D. Chen
 for helpful discussions. This work was supported by the ERC Advanced Grant MiCE. The authors would also like to acknowledge the use of the Advanced Research Computing (ARC)
  in carrying out this work.


\bibliography{refe.bib}

\providecommand{\noopsort}[1]{}\providecommand{\singleletter}[1]{#1}%
\begin{thebibliography}{34}%
\makeatletter
\providecommand \@ifxundefined [1]{%
 \@ifx{#1\undefined}
}%
\providecommand \@ifnum [1]{%
 \ifnum #1\expandafter \@firstoftwo
 \else \expandafter \@secondoftwo
 \fi
}%
\providecommand \@ifx [1]{%
 \ifx #1\expandafter \@firstoftwo
 \else \expandafter \@secondoftwo
 \fi
}%
\providecommand \natexlab [1]{#1}%
\providecommand \enquote  [1]{``#1''}%
\providecommand \bibnamefont  [1]{#1}%
\providecommand \bibfnamefont [1]{#1}%
\providecommand \citenamefont [1]{#1}%
\providecommand \href@noop [0]{\@secondoftwo}%
\providecommand \href [0]{\begingroup \@sanitize@url \@href}%
\providecommand \@href[1]{\@@startlink{#1}\@@href}%
\providecommand \@@href[1]{\endgroup#1\@@endlink}%
\providecommand \@sanitize@url [0]{\catcode `\\12\catcode `\$12\catcode
  `\&12\catcode `\#12\catcode `\^12\catcode `\_12\catcode `\%12\relax}%
\providecommand \@@startlink[1]{}%
\providecommand \@@endlink[0]{}%
\providecommand \url  [0]{\begingroup\@sanitize@url \@url }%
\providecommand \@url [1]{\endgroup\@href {#1}{\urlprefix }}%
\providecommand \urlprefix  [0]{URL }%
\providecommand \Eprint [0]{\href }%
\providecommand \doibase [0]{http://dx.doi.org/}%
\providecommand \selectlanguage [0]{\@gobble}%
\providecommand \bibinfo  [0]{\@secondoftwo}%
\providecommand \bibfield  [0]{\@secondoftwo}%
\providecommand \translation [1]{[#1]}%
\providecommand \BibitemOpen [0]{}%
\providecommand \bibitemStop [0]{}%
\providecommand \bibitemNoStop [0]{.\EOS\space}%
\providecommand \EOS [0]{\spacefactor3000\relax}%
\providecommand \BibitemShut  [1]{\csname bibitem#1\endcsname}%
\let\auto@bib@innerbib\@empty
\bibitem [{\citenamefont {Sanchez}\ \emph {et~al.}(2012)\citenamefont
  {Sanchez}, \citenamefont {Chen}, \citenamefont {DeCamp}, \citenamefont
  {Heymann},\ and\ \citenamefont {Dogic}}]{dogic2012}%
  \BibitemOpen
  \bibfield  {author} {\bibinfo {author} {\bibfnamefont {T.}~\bibnamefont
  {Sanchez}}, \bibinfo {author} {\bibfnamefont {D.~T.~N.}\ \bibnamefont
  {Chen}}, \bibinfo {author} {\bibfnamefont {S.~J.}\ \bibnamefont {DeCamp}},
  \bibinfo {author} {\bibfnamefont {M.}~\bibnamefont {Heymann}}, \ and\
  \bibinfo {author} {\bibfnamefont {Z.}~\bibnamefont {Dogic}},\ }\href
  {\doibase 10.1038/nature11591} {\bibfield  {journal} {\bibinfo  {journal}
  {Nature}\ }\textbf {\bibinfo {volume} {491}},\ \bibinfo {pages} {431}
  (\bibinfo {year} {2012})}\BibitemShut {NoStop}%
\bibitem [{\citenamefont {Ramaswamy}(2010)}]{Sriram2010}%
  \BibitemOpen
  \bibfield  {author} {\bibinfo {author} {\bibfnamefont {S.}~\bibnamefont
  {Ramaswamy}},\ }\href {\doibase 10.1146/annurev-conmatphys-070909-104101}
  {\bibfield  {journal} {\bibinfo  {journal} {Annu. Rev. Cond. Mat. Phys.}\
  }\textbf {\bibinfo {volume} {1}},\ \bibinfo {pages} {323} (\bibinfo {year}
  {2010})}\BibitemShut {NoStop}%
\bibitem [{\citenamefont {Marchetti}\ \emph {et~al.}(2013)\citenamefont
  {Marchetti}, \citenamefont {Joanny}, \citenamefont {Ramaswamy}, \citenamefont
  {Liverpool}, \citenamefont {Prost}, \citenamefont {Rao},\ and\ \citenamefont
  {Simha}}]{Marchetti2013}%
  \BibitemOpen
  \bibfield  {author} {\bibinfo {author} {\bibfnamefont {M.~C.}\ \bibnamefont
  {Marchetti}}, \bibinfo {author} {\bibfnamefont {J.~F.}\ \bibnamefont
  {Joanny}}, \bibinfo {author} {\bibfnamefont {S.}~\bibnamefont {Ramaswamy}},
  \bibinfo {author} {\bibfnamefont {T.~B.}\ \bibnamefont {Liverpool}}, \bibinfo
  {author} {\bibfnamefont {J.}~\bibnamefont {Prost}}, \bibinfo {author}
  {\bibfnamefont {M.}~\bibnamefont {Rao}}, \ and\ \bibinfo {author}
  {\bibfnamefont {R.~A.}\ \bibnamefont {Simha}},\ }\href {\doibase
  10.1103/RevModPhys.85.1143} {\bibfield  {journal} {\bibinfo  {journal} {Rev.
  Mod. Phys.}\ }\textbf {\bibinfo {volume} {85}},\ \bibinfo {pages} {1143}
  (\bibinfo {year} {2013})}\BibitemShut {NoStop}%
\bibitem [{\citenamefont {Dunkel}\ \emph {et~al.}(2013)\citenamefont {Dunkel},
  \citenamefont {Heidenreich}, \citenamefont {Drescher}, \citenamefont
  {Wensink}, \citenamefont {B\"ar},\ and\ \citenamefont
  {Goldstein}}]{Jorn2013}%
  \BibitemOpen
  \bibfield  {author} {\bibinfo {author} {\bibfnamefont {J.}~\bibnamefont
  {Dunkel}}, \bibinfo {author} {\bibfnamefont {S.}~\bibnamefont {Heidenreich}},
  \bibinfo {author} {\bibfnamefont {K.}~\bibnamefont {Drescher}}, \bibinfo
  {author} {\bibfnamefont {H.~H.}\ \bibnamefont {Wensink}}, \bibinfo {author}
  {\bibfnamefont {M.}~\bibnamefont {B\"ar}}, \ and\ \bibinfo {author}
  {\bibfnamefont {R.~E.}\ \bibnamefont {Goldstein}},\ }\href {\doibase
  10.1103/PhysRevLett.110.228102} {\bibfield  {journal} {\bibinfo  {journal}
  {Phys. Rev. Lett.}\ }\textbf {\bibinfo {volume} {110}},\ \bibinfo {pages}
  {228102} (\bibinfo {year} {2013})}\BibitemShut {NoStop}%
\bibitem [{\citenamefont {Sumino}\ \emph {et~al.}(2012)\citenamefont {Sumino},
  \citenamefont {Nagai}, \citenamefont {Shitaka}, \citenamefont {Tanaka},
  \citenamefont {Yoshikawa}, \citenamefont {Chate},\ and\ \citenamefont
  {Oiwa}}]{Chate2012}%
  \BibitemOpen
  \bibfield  {author} {\bibinfo {author} {\bibfnamefont {Y.}~\bibnamefont
  {Sumino}}, \bibinfo {author} {\bibfnamefont {K.~H.}\ \bibnamefont {Nagai}},
  \bibinfo {author} {\bibfnamefont {Y.}~\bibnamefont {Shitaka}}, \bibinfo
  {author} {\bibfnamefont {D.}~\bibnamefont {Tanaka}}, \bibinfo {author}
  {\bibfnamefont {K.}~\bibnamefont {Yoshikawa}}, \bibinfo {author}
  {\bibfnamefont {H.}~\bibnamefont {Chate}}, \ and\ \bibinfo {author}
  {\bibfnamefont {K.}~\bibnamefont {Oiwa}},\ }\href {\doibase
  10.1038/nature10874} {\bibfield  {journal} {\bibinfo  {journal} {Nature}\
  }\textbf {\bibinfo {volume} {483}},\ \bibinfo {pages} {448} (\bibinfo {year}
  {2012})}\BibitemShut {NoStop}%
\bibitem [{\citenamefont {Cavagna}\ \emph {et~al.}(2010)\citenamefont
  {Cavagna}, \citenamefont {Cimarelli}, \citenamefont {Giardina}, \citenamefont
  {Parisi}, \citenamefont {Santagati}, \citenamefont {Stefanini},\ and\
  \citenamefont {Viale}}]{Cavagna2010}%
  \BibitemOpen
  \bibfield  {author} {\bibinfo {author} {\bibfnamefont {A.}~\bibnamefont
  {Cavagna}}, \bibinfo {author} {\bibfnamefont {A.}~\bibnamefont {Cimarelli}},
  \bibinfo {author} {\bibfnamefont {I.}~\bibnamefont {Giardina}}, \bibinfo
  {author} {\bibfnamefont {G.}~\bibnamefont {Parisi}}, \bibinfo {author}
  {\bibfnamefont {R.}~\bibnamefont {Santagati}}, \bibinfo {author}
  {\bibfnamefont {F.}~\bibnamefont {Stefanini}}, \ and\ \bibinfo {author}
  {\bibfnamefont {M.}~\bibnamefont {Viale}},\ }\href {\doibase
  10.1073/pnas.1005766107} {\bibfield  {journal} {\bibinfo  {journal} {PNAS}\
  }\textbf {\bibinfo {volume} {107}},\ \bibinfo {pages} {11865} (\bibinfo
  {year} {2010})}\BibitemShut {NoStop}%
\bibitem [{\citenamefont {Wensink}\ \emph {et~al.}(2012)\citenamefont
  {Wensink}, \citenamefont {Dunkel}, \citenamefont {Heidenreich}, \citenamefont
  {Drescher}, \citenamefont {Goldstein}, \citenamefont {Lowen},\ and\
  \citenamefont {Yeomans}}]{julia2012}%
  \BibitemOpen
  \bibfield  {author} {\bibinfo {author} {\bibfnamefont {H.~H.}\ \bibnamefont
  {Wensink}}, \bibinfo {author} {\bibfnamefont {J.}~\bibnamefont {Dunkel}},
  \bibinfo {author} {\bibfnamefont {S.}~\bibnamefont {Heidenreich}}, \bibinfo
  {author} {\bibfnamefont {K.}~\bibnamefont {Drescher}}, \bibinfo {author}
  {\bibfnamefont {R.~E.}\ \bibnamefont {Goldstein}}, \bibinfo {author}
  {\bibfnamefont {H.}~\bibnamefont {Lowen}}, \ and\ \bibinfo {author}
  {\bibfnamefont {J.~M.}\ \bibnamefont {Yeomans}},\ }\href {\doibase
  10.1073/pnas.1202032109} {\bibfield  {journal} {\bibinfo  {journal} {PNAS}\
  }\textbf {\bibinfo {volume} {109}},\ \bibinfo {pages} {14308} (\bibinfo
  {year} {2012})}\BibitemShut {NoStop}%
\bibitem [{\citenamefont {A.~Simha}\ and\ \citenamefont
  {Ramaswamy}(2002)}]{Sriram2002}%
  \BibitemOpen
  \bibfield  {author} {\bibinfo {author} {\bibfnamefont {R.}~\bibnamefont
  {A.~Simha}}\ and\ \bibinfo {author} {\bibfnamefont {S.}~\bibnamefont
  {Ramaswamy}},\ }\href {\doibase 10.1103/PhysRevLett.89.058101} {\bibfield
  {journal} {\bibinfo  {journal} {Phys. Rev. Lett.}\ }\textbf {\bibinfo
  {volume} {89}},\ \bibinfo {pages} {058101} (\bibinfo {year}
  {2002})}\BibitemShut {NoStop}%
\bibitem [{\citenamefont {Voituriez}\ \emph {et~al.}(2005)\citenamefont
  {Voituriez}, \citenamefont {Joanny},\ and\ \citenamefont
  {Prost}}]{Joanny2005}%
  \BibitemOpen
  \bibfield  {author} {\bibinfo {author} {\bibfnamefont {R.}~\bibnamefont
  {Voituriez}}, \bibinfo {author} {\bibfnamefont {J.~F.}\ \bibnamefont
  {Joanny}}, \ and\ \bibinfo {author} {\bibfnamefont {J.}~\bibnamefont
  {Prost}},\ }\href {\doibase 10.1209/epl/i2004-10501-2} {\bibfield  {journal}
  {\bibinfo  {journal} {Europhys. Lett.}\ }\textbf {\bibinfo {volume} {70}},\
  \bibinfo {pages} {404} (\bibinfo {year} {2005})}\BibitemShut {NoStop}%
\bibitem [{\citenamefont {Ramaswamy}\ and\ \citenamefont
  {Rao}(2007)}]{Madan2007}%
  \BibitemOpen
  \bibfield  {author} {\bibinfo {author} {\bibfnamefont {S.}~\bibnamefont
  {Ramaswamy}}\ and\ \bibinfo {author} {\bibfnamefont {M.}~\bibnamefont
  {Rao}},\ }\href {\doibase 10.1088/1367-2630/9/11/423} {\bibfield  {journal}
  {\bibinfo  {journal} {New J. Phys.}\ }\textbf {\bibinfo {volume} {9}},\
  \bibinfo {pages} {423} (\bibinfo {year} {2007})}\BibitemShut {NoStop}%
\bibitem [{\citenamefont {Marenduzzo}\ \emph
  {et~al.}(2007{\natexlab{a}})\citenamefont {Marenduzzo}, \citenamefont
  {Orlandini}, \citenamefont {Cates},\ and\ \citenamefont
  {Yeomans}}]{Davide2007}%
  \BibitemOpen
  \bibfield  {author} {\bibinfo {author} {\bibfnamefont {D.}~\bibnamefont
  {Marenduzzo}}, \bibinfo {author} {\bibfnamefont {E.}~\bibnamefont
  {Orlandini}}, \bibinfo {author} {\bibfnamefont {M.~E.}\ \bibnamefont
  {Cates}}, \ and\ \bibinfo {author} {\bibfnamefont {J.~M.}\ \bibnamefont
  {Yeomans}},\ }\href {\doibase 10.1103/PhysRevE.76.031921} {\bibfield
  {journal} {\bibinfo  {journal} {Phys. Rev. E}\ }\textbf {\bibinfo {volume}
  {76}},\ \bibinfo {pages} {031921} (\bibinfo {year}
  {2007}{\natexlab{a}})}\BibitemShut {NoStop}%
\bibitem [{\citenamefont {Edwards}\ and\ \citenamefont
  {Yeomans}(2009)}]{Scott2009}%
  \BibitemOpen
  \bibfield  {author} {\bibinfo {author} {\bibfnamefont {S.}~\bibnamefont
  {Edwards}}\ and\ \bibinfo {author} {\bibfnamefont {J.~M.}\ \bibnamefont
  {Yeomans}},\ }\href {\doibase 10.1209/0295-5075/85/18008} {\bibfield
  {journal} {\bibinfo  {journal} {Europhys. Lett.}\ }\textbf {\bibinfo {volume}
  {85}},\ \bibinfo {pages} {18008} (\bibinfo {year} {2009})}\BibitemShut
  {NoStop}%
\bibitem [{\citenamefont {Giomi}\ \emph {et~al.}(2011)\citenamefont {Giomi},
  \citenamefont {Mahadevan}, \citenamefont {Chakraborty},\ and\ \citenamefont
  {Hagan}}]{Mahadevan2011}%
  \BibitemOpen
  \bibfield  {author} {\bibinfo {author} {\bibfnamefont {L.}~\bibnamefont
  {Giomi}}, \bibinfo {author} {\bibfnamefont {L.}~\bibnamefont {Mahadevan}},
  \bibinfo {author} {\bibfnamefont {B.}~\bibnamefont {Chakraborty}}, \ and\
  \bibinfo {author} {\bibfnamefont {M.~F.}\ \bibnamefont {Hagan}},\ }\href
  {\doibase 10.1103/PhysRevLett.106.218101} {\bibfield  {journal} {\bibinfo
  {journal} {Phys. Rev. Lett.}\ }\textbf {\bibinfo {volume} {106}},\ \bibinfo
  {pages} {218101} (\bibinfo {year} {2011})}\BibitemShut {NoStop}%
\bibitem [{\citenamefont {Thampi}\ \emph {et~al.}(2013)\citenamefont {Thampi},
  \citenamefont {Golestanian},\ and\ \citenamefont {Yeomans}}]{ourprl2013}%
  \BibitemOpen
  \bibfield  {author} {\bibinfo {author} {\bibfnamefont {S.~P.}\ \bibnamefont
  {Thampi}}, \bibinfo {author} {\bibfnamefont {R.}~\bibnamefont {Golestanian}},
  \ and\ \bibinfo {author} {\bibfnamefont {J.~M.}\ \bibnamefont {Yeomans}},\
  }\href {\doibase 10.1103/PhysRevLett.111.118101} {\bibfield  {journal}
  {\bibinfo  {journal} {Phys. Rev. Lett.}\ }\textbf {\bibinfo {volume} {111}},\
  \bibinfo {pages} {118101} (\bibinfo {year} {2013})}\BibitemShut {NoStop}%
\bibitem [{\citenamefont {Thampi}\ \emph {et~al.}(2014)\citenamefont {Thampi},
  \citenamefont {Golestanian},\ and\ \citenamefont {Yeomans}}]{ourepl2014}%
  \BibitemOpen
  \bibfield  {author} {\bibinfo {author} {\bibfnamefont {S.~P.}\ \bibnamefont
  {Thampi}}, \bibinfo {author} {\bibfnamefont {R.}~\bibnamefont {Golestanian}},
  \ and\ \bibinfo {author} {\bibfnamefont {J.~M.}\ \bibnamefont {Yeomans}},\
  }\href {\doibase 10.1209/0295-5075/105/18001} {\bibfield  {journal} {\bibinfo
   {journal} {Europhys. Lett.}\ }\textbf {\bibinfo {volume} {105}},\ \bibinfo
  {pages} {18001} (\bibinfo {year} {2014})}\BibitemShut {NoStop}%
\bibitem [{\citenamefont {Giomi}\ \emph {et~al.}(2013)\citenamefont {Giomi},
  \citenamefont {Bowick}, \citenamefont {Ma},\ and\ \citenamefont
  {Marchetti}}]{Giomi2013}%
  \BibitemOpen
  \bibfield  {author} {\bibinfo {author} {\bibfnamefont {L.}~\bibnamefont
  {Giomi}}, \bibinfo {author} {\bibfnamefont {M.~J.}\ \bibnamefont {Bowick}},
  \bibinfo {author} {\bibfnamefont {X.}~\bibnamefont {Ma}}, \ and\ \bibinfo
  {author} {\bibfnamefont {M.~C.}\ \bibnamefont {Marchetti}},\ }\href {\doibase
  10.1103/PhysRevLett.110.228101} {\bibfield  {journal} {\bibinfo  {journal}
  {Phys. Rev. Lett.}\ }\textbf {\bibinfo {volume} {110}},\ \bibinfo {pages}
  {228101} (\bibinfo {year} {2013})}\BibitemShut {NoStop}%
\bibitem [{\citenamefont {Zhou}\ \emph {et~al.}(2014)\citenamefont {Zhou},
  \citenamefont {Sokolov}, \citenamefont {Lavrentovich},\ and\ \citenamefont
  {Aranson}}]{Igor2013}%
  \BibitemOpen
  \bibfield  {author} {\bibinfo {author} {\bibfnamefont {S.}~\bibnamefont
  {Zhou}}, \bibinfo {author} {\bibfnamefont {A.}~\bibnamefont {Sokolov}},
  \bibinfo {author} {\bibfnamefont {O.~D.}\ \bibnamefont {Lavrentovich}}, \
  and\ \bibinfo {author} {\bibfnamefont {I.~S.}\ \bibnamefont {Aranson}},\
  }\href {\doibase 10.1073/pnas.1321926111} {\bibfield  {journal} {\bibinfo
  {journal} {PNAS}\ }\textbf {\bibinfo {volume} {111}},\ \bibinfo {pages}
  {1265} (\bibinfo {year} {2014})}\BibitemShut {NoStop}%
\bibitem [{\citenamefont {Narayan}\ \emph {et~al.}(2007)\citenamefont
  {Narayan}, \citenamefont {Ramaswamy},\ and\ \citenamefont
  {Menon}}]{Narayan2007}%
  \BibitemOpen
  \bibfield  {author} {\bibinfo {author} {\bibfnamefont {V.}~\bibnamefont
  {Narayan}}, \bibinfo {author} {\bibfnamefont {S.}~\bibnamefont {Ramaswamy}},
  \ and\ \bibinfo {author} {\bibfnamefont {N.}~\bibnamefont {Menon}},\ }\href
  {\doibase 10.1126/science.1140414} {\bibfield  {journal} {\bibinfo  {journal}
  {Science}\ }\textbf {\bibinfo {volume} {317}},\ \bibinfo {pages} {105}
  (\bibinfo {year} {2007})}\BibitemShut {NoStop}%
\bibitem [{\citenamefont {Shi}\ and\ \citenamefont {Ma}(2013)}]{Shi2013}%
  \BibitemOpen
  \bibfield  {author} {\bibinfo {author} {\bibfnamefont {X.}~\bibnamefont
  {Shi}}\ and\ \bibinfo {author} {\bibfnamefont {Y.}~\bibnamefont {Ma}},\
  }\href {\doibase 10.1038/ncomms4013} {\bibfield  {journal} {\bibinfo
  {journal} {Nat. Commun.}\ }\textbf {\bibinfo {volume} {4}},\ \bibinfo {pages}
  {3013} (\bibinfo {year} {2013})}\BibitemShut {NoStop}%
\bibitem [{\citenamefont {Beris}\ and\ \citenamefont
  {Edwards}(1994)}]{Berisbook}%
  \BibitemOpen
  \bibfield  {author} {\bibinfo {author} {\bibfnamefont {A.~N.}\ \bibnamefont
  {Beris}}\ and\ \bibinfo {author} {\bibfnamefont {B.~J.}\ \bibnamefont
  {Edwards}},\ }\href@noop {} {\emph {\bibinfo {title} {Thermodynamics of
  Flowing Systems}}}\ (\bibinfo  {publisher} {Oxford University Press},\
  \bibinfo {year} {1994})\BibitemShut {NoStop}%
\bibitem [{\citenamefont {de~Gennes}\ and\ \citenamefont
  {Prost}(1995)}]{DeGennesBook}%
  \BibitemOpen
  \bibfield  {author} {\bibinfo {author} {\bibfnamefont {P.~G.}\ \bibnamefont
  {de~Gennes}}\ and\ \bibinfo {author} {\bibfnamefont {J.}~\bibnamefont
  {Prost}},\ }\href@noop {} {\emph {\bibinfo {title} {The Physics of Liquid
  Crystals}}}\ (\bibinfo  {publisher} {Oxford University Press},\ \bibinfo
  {year} {1995})\BibitemShut {NoStop}%
\bibitem [{\citenamefont {T\'oth}\ \emph {et~al.}(2002)\citenamefont {T\'oth},
  \citenamefont {Denniston},\ and\ \citenamefont {Yeomans}}]{Julia2002}%
  \BibitemOpen
  \bibfield  {author} {\bibinfo {author} {\bibfnamefont {G.}~\bibnamefont
  {T\'oth}}, \bibinfo {author} {\bibfnamefont {C.}~\bibnamefont {Denniston}}, \
  and\ \bibinfo {author} {\bibfnamefont {J.~M.}\ \bibnamefont {Yeomans}},\
  }\href {\doibase 10.1103/PhysRevLett.88.105504} {\bibfield  {journal}
  {\bibinfo  {journal} {Phys. Rev. Lett.}\ }\textbf {\bibinfo {volume} {88}},\
  \bibinfo {pages} {105504} (\bibinfo {year} {2002})}\BibitemShut {NoStop}%
\bibitem [{\citenamefont {Denniston}\ \emph {et~al.}(2001)\citenamefont
  {Denniston}, \citenamefont {Orlandini},\ and\ \citenamefont
  {Yeomans}}]{Denniston2001}%
  \BibitemOpen
  \bibfield  {author} {\bibinfo {author} {\bibfnamefont {C.}~\bibnamefont
  {Denniston}}, \bibinfo {author} {\bibfnamefont {E.}~\bibnamefont
  {Orlandini}}, \ and\ \bibinfo {author} {\bibfnamefont {J.~M.}\ \bibnamefont
  {Yeomans}},\ }\href {\doibase 10.1103/PhysRevE.63.056702} {\bibfield
  {journal} {\bibinfo  {journal} {Phys. Rev. E}\ }\textbf {\bibinfo {volume}
  {63}},\ \bibinfo {pages} {056702} (\bibinfo {year} {2001})}\BibitemShut
  {NoStop}%
\bibitem [{\citenamefont {Denniston}\ \emph {et~al.}(2004)\citenamefont
  {Denniston}, \citenamefont {Marenduzzo}, \citenamefont {Orlandini},\ and\
  \citenamefont {Yeomans}}]{Denniston2004}%
  \BibitemOpen
  \bibfield  {author} {\bibinfo {author} {\bibfnamefont {C.}~\bibnamefont
  {Denniston}}, \bibinfo {author} {\bibfnamefont {D.}~\bibnamefont
  {Marenduzzo}}, \bibinfo {author} {\bibfnamefont {E.}~\bibnamefont
  {Orlandini}}, \ and\ \bibinfo {author} {\bibfnamefont {J.~M.}\ \bibnamefont
  {Yeomans}},\ }\href {\doibase 10.1098/rsta.2004.1416} {\bibfield  {journal}
  {\bibinfo  {journal} {Phil. Trans. R. Soc. Lond. A}\ }\textbf {\bibinfo
  {volume} {362}},\ \bibinfo {pages} {1745} (\bibinfo {year}
  {2004})}\BibitemShut {NoStop}%
\bibitem [{\citenamefont {Cates}\ \emph {et~al.}(2008)\citenamefont {Cates},
  \citenamefont {Fielding}, \citenamefont {Marenduzzo}, \citenamefont
  {Orlandini},\ and\ \citenamefont {Yeomans}}]{Cates2008}%
  \BibitemOpen
  \bibfield  {author} {\bibinfo {author} {\bibfnamefont {M.~E.}\ \bibnamefont
  {Cates}}, \bibinfo {author} {\bibfnamefont {S.~M.}\ \bibnamefont {Fielding}},
  \bibinfo {author} {\bibfnamefont {D.}~\bibnamefont {Marenduzzo}}, \bibinfo
  {author} {\bibfnamefont {E.}~\bibnamefont {Orlandini}}, \ and\ \bibinfo
  {author} {\bibfnamefont {J.~M.}\ \bibnamefont {Yeomans}},\ }\href {\doibase
  10.1103/PhysRevLett.101.068102} {\bibfield  {journal} {\bibinfo  {journal}
  {Phys. Rev. Lett.}\ }\textbf {\bibinfo {volume} {101}},\ \bibinfo {pages}
  {068102} (\bibinfo {year} {2008})}\BibitemShut {NoStop}%
\bibitem [{\citenamefont {Orlandini}\ \emph {et~al.}(2008)\citenamefont
  {Orlandini}, \citenamefont {Cates}, \citenamefont {Marenduzzo}, \citenamefont
  {Tubiana},\ and\ \citenamefont {Yeomans}}]{Orlandini2008}%
  \BibitemOpen
  \bibfield  {author} {\bibinfo {author} {\bibfnamefont {E.}~\bibnamefont
  {Orlandini}}, \bibinfo {author} {\bibfnamefont {M.~E.}\ \bibnamefont
  {Cates}}, \bibinfo {author} {\bibfnamefont {D.}~\bibnamefont {Marenduzzo}},
  \bibinfo {author} {\bibfnamefont {L.}~\bibnamefont {Tubiana}}, \ and\
  \bibinfo {author} {\bibfnamefont {J.~M.}\ \bibnamefont {Yeomans}},\ }\href
  {\doibase 10.1080/15421400802430117} {\bibfield  {journal} {\bibinfo
  {journal} {Mol. Cryst. and Liq. Cryst.}\ }\textbf {\bibinfo {volume} {494}},\
  \bibinfo {pages} {293} (\bibinfo {year} {2008})}\BibitemShut {NoStop}%
\bibitem [{\citenamefont {Henrich}\ \emph {et~al.}(2010)\citenamefont
  {Henrich}, \citenamefont {Stratford}, \citenamefont {Marenduzzo},\ and\
  \citenamefont {Cates}}]{Henrich2010}%
  \BibitemOpen
  \bibfield  {author} {\bibinfo {author} {\bibfnamefont {O.}~\bibnamefont
  {Henrich}}, \bibinfo {author} {\bibfnamefont {K.}~\bibnamefont {Stratford}},
  \bibinfo {author} {\bibfnamefont {D.}~\bibnamefont {Marenduzzo}}, \ and\
  \bibinfo {author} {\bibfnamefont {M.~E.}\ \bibnamefont {Cates}},\ }\href
  {\doibase 10.1073/pnas.1004269107} {\bibfield  {journal} {\bibinfo  {journal}
  {PNAS}\ }\textbf {\bibinfo {volume} {107}},\ \bibinfo {pages} {13212}
  (\bibinfo {year} {2010})}\BibitemShut {NoStop}%
\bibitem [{\citenamefont {Fielding}\ \emph {et~al.}(2011)\citenamefont
  {Fielding}, \citenamefont {Marenduzzo},\ and\ \citenamefont
  {Cates}}]{Suzanne2011}%
  \BibitemOpen
  \bibfield  {author} {\bibinfo {author} {\bibfnamefont {S.~M.}\ \bibnamefont
  {Fielding}}, \bibinfo {author} {\bibfnamefont {D.}~\bibnamefont
  {Marenduzzo}}, \ and\ \bibinfo {author} {\bibfnamefont {M.~E.}\ \bibnamefont
  {Cates}},\ }\href {\doibase 10.1103/PhysRevE.83.041910} {\bibfield  {journal}
  {\bibinfo  {journal} {Phys. Rev. E}\ }\textbf {\bibinfo {volume} {83}},\
  \bibinfo {pages} {041910} (\bibinfo {year} {2011})}\BibitemShut {NoStop}%
\bibitem [{\citenamefont {Ravnik}\ and\ \citenamefont
  {Yeomans}(2013)}]{Miha2013}%
  \BibitemOpen
  \bibfield  {author} {\bibinfo {author} {\bibfnamefont {M.}~\bibnamefont
  {Ravnik}}\ and\ \bibinfo {author} {\bibfnamefont {J.~M.}\ \bibnamefont
  {Yeomans}},\ }\href {\doibase 10.1103/PhysRevLett.110.026001} {\bibfield
  {journal} {\bibinfo  {journal} {Phys. Rev. Lett.}\ }\textbf {\bibinfo
  {volume} {110}},\ \bibinfo {pages} {026001} (\bibinfo {year}
  {2013})}\BibitemShut {NoStop}%
\bibitem [{\citenamefont {Cates}\ \emph {et~al.}(2009)\citenamefont {Cates},
  \citenamefont {Henrich}, \citenamefont {Marenduzzo},\ and\ \citenamefont
  {Stratford}}]{Cates2009}%
  \BibitemOpen
  \bibfield  {author} {\bibinfo {author} {\bibfnamefont {M.~E.}\ \bibnamefont
  {Cates}}, \bibinfo {author} {\bibfnamefont {O.}~\bibnamefont {Henrich}},
  \bibinfo {author} {\bibfnamefont {D.}~\bibnamefont {Marenduzzo}}, \ and\
  \bibinfo {author} {\bibfnamefont {K.}~\bibnamefont {Stratford}},\ }\href
  {\doibase 10.1039/B908659P} {\bibfield  {journal} {\bibinfo  {journal} {Soft
  Matter}\ }\textbf {\bibinfo {volume} {5}},\ \bibinfo {pages} {3791} (\bibinfo
  {year} {2009})}\BibitemShut {NoStop}%
\bibitem [{\citenamefont {Marenduzzo}\ \emph
  {et~al.}(2007{\natexlab{b}})\citenamefont {Marenduzzo}, \citenamefont
  {Orlandini},\ and\ \citenamefont {Yeomans}}]{Davide2007B}%
  \BibitemOpen
  \bibfield  {author} {\bibinfo {author} {\bibfnamefont {D.}~\bibnamefont
  {Marenduzzo}}, \bibinfo {author} {\bibfnamefont {E.}~\bibnamefont
  {Orlandini}}, \ and\ \bibinfo {author} {\bibfnamefont {J.~M.}\ \bibnamefont
  {Yeomans}},\ }\href {\doibase 10.1103/PhysRevLett.98.118102} {\bibfield
  {journal} {\bibinfo  {journal} {Phys. Rev. Lett.}\ }\textbf {\bibinfo
  {volume} {98}},\ \bibinfo {pages} {118102} (\bibinfo {year}
  {2007}{\natexlab{b}})}\BibitemShut {NoStop}%
\bibitem [{\citenamefont {Rossen}\ \emph {et~al.}(2013)\citenamefont {Rossen},
  \citenamefont {Jensen},\ and\ \citenamefont {Oddershede}}]{Lenepreprint}%
  \BibitemOpen
  \bibfield  {author} {\bibinfo {author} {\bibfnamefont {N.~S.}\ \bibnamefont
  {Rossen}}, \bibinfo {author} {\bibfnamefont {M.~H.}\ \bibnamefont {Jensen}},
  \ and\ \bibinfo {author} {\bibfnamefont {L.~B.}\ \bibnamefont {Oddershede}},\
  }\href@noop {} {\enquote {\bibinfo {title} {Long range ordered vorticity
  patterns induced by cell division},}\ }\bibinfo {howpublished} {Private
  Communication} (\bibinfo {year} {2013})\BibitemShut {NoStop}%
\bibitem [{\citenamefont {Woodhouse}\ and\ \citenamefont
  {Goldstein}(2012)}]{Goldstein2012}%
  \BibitemOpen
  \bibfield  {author} {\bibinfo {author} {\bibfnamefont {F.~G.}\ \bibnamefont
  {Woodhouse}}\ and\ \bibinfo {author} {\bibfnamefont {R.~E.}\ \bibnamefont
  {Goldstein}},\ }\href {\doibase 10.1103/PhysRevLett.109.168105} {\bibfield
  {journal} {\bibinfo  {journal} {Phys. Rev. Lett.}\ }\textbf {\bibinfo
  {volume} {109}},\ \bibinfo {pages} {168105} (\bibinfo {year}
  {2012})}\BibitemShut {NoStop}%
\bibitem [{\citenamefont {Sokolov}\ and\ \citenamefont
  {Aranson}(2012)}]{Aranson2012}%
  \BibitemOpen
  \bibfield  {author} {\bibinfo {author} {\bibfnamefont {A.}~\bibnamefont
  {Sokolov}}\ and\ \bibinfo {author} {\bibfnamefont {I.~S.}\ \bibnamefont
  {Aranson}},\ }\href {\doibase 10.1103/PhysRevLett.109.248109} {\bibfield
  {journal} {\bibinfo  {journal} {Phys. Rev. Lett.}\ }\textbf {\bibinfo
  {volume} {109}},\ \bibinfo {pages} {248109} (\bibinfo {year}
  {2012})}\BibitemShut {NoStop}%
\end{thebibliography}%

\end{document}